\documentclass{vgtc}                          

\ifpdf
  \pdfoutput=1\relax                   
  \usepackage{graphicx}                
  \DeclareGraphicsExtensions{.pdf,.png,.jpg,.jpeg} 
\else
  \usepackage{graphicx}                
  \DeclareGraphicsExtensions{.eps}     
\fi%

\graphicspath{{figures/}{pictures/}{images/}{./}} 

\usepackage{microtype}                 
\PassOptionsToPackage{warn}{textcomp}  
\usepackage{textcomp}                  
\usepackage{mathptmx}                  
\usepackage{times}                     
\usepackage{cite}                      
\usepackage{tabu}                      
\usepackage{booktabs}                  

\usepackage{enumitem}

\onlineid{0}

\vgtccategory{Research}

\vgtcinsertpkg

\title{Reflections and Considerations on \\Running Creative Visualization Learning Activities}

\newcommand{\super}[1]{\ensuremath{^{{\scriptsize #1}}}}
\newcommand{\mthanks}[2]{\super{#1}#2}

\author{
Jonathan C. Roberts\thanks{j.c.roberts@bangor.ac.uk,
\raggedright 
\mthanks{2}{bbach@ed.ac.uk}, 
\mthanks{3}{magdalena.boucher@fhstp.ac.at},
\mthanks{4}{fanny@cs.toronto.edu},
\mthanks{5}{diehl@ifi.uzh.ch}, 
\mthanks{6}{uhinrich@ed.ac.uk},
\mthanks{7}{samuel.huron@telecom-paris.fr}, 
\mthanks{8}{andy@visualisingdata.com},
\mthanks{9}{soekn@itu.dk}, 
\mthanks{10}{imeirelles@ocadu.ca}, 
\mthanks{11}{rebecca.noonan@mycit.ie},
\mthanks{12}{pelchmann@cs.uni-koeln.de},
\mthanks{13}{fateme.rajabiyazdi@carleton.ca},
\mthanks{14}{cstoiber@fhstp.ac.at}
}\\
\scriptsize{Bangor University}
\and
Benjamin Bach\super{2}\\ \scriptsize{University of Edinburgh}
\and Magdalena Boucher\super{3}\\ \scriptsize{St. Pölten University of Applied Sciences}
\and Fanny Chevalier\super{4}\\\scriptsize{University of Toronto}\vspace{1mm}
\and Alexandra Diehl\super{5}\\\scriptsize{University of Zurich}
\and Uta Hinrichs\super{6}\\\scriptsize{University of Edinburgh}
\and Samuel Huron\super{7}\\\scriptsize{\centering Telecom Paris \& Institut Polytechnique de Paris}
\and Andy Kirk\super{8}\\ \scriptsize{Visualising Data Ltd}\vspace{1mm}
\and Søren Knudsen\super{9}\\\scriptsize{IT University of Copenhagen}
\and Isabel Meirelles\super{10}\\\scriptsize{OCAD University}
\and Rebecca Noonan\super{11}\\\scriptsize{Munster Technological University}
\and Laura Pelchmann\super{12}\\\scriptsize{University of Cologne}\vspace{1mm}
\and Fateme Rajabiyazdi\super{13}\\\scriptsize{Carleton University}
\and Christina Stoiber\super{14}\\ \scriptsize{St. Pölten University of Applied Sciences}
}

\abstract{This paper draws together nine strategies for creative visualization activities. Teaching visualization often involves running learning activities where students perform tasks that directly support one or more topics that the teacher wishes to address in the lesson. As a group of educators and researchers in visualization, we reflect on our learning experiences. Our activities and experiences range from dividing the tasks into smaller parts, considering different learning materials, to encouraging debate. With this paper, our hope is that we can encourage, inspire, and guide other educators with visualization activities. Our reflections provide an initial starting point of methods and strategies to craft creative visualisation learning activities, and provide a foundation for developing best practices in visualization education. %
} 


\CCScatlist{
  \CCScatTwelve{Human-centered computing}{Visu\-al\-iza\-tion}{}{};
  \CCScatTwelve{Applied Computing}{Education}{}{};
}

\begin{document}

\firstsection{Introduction}
\maketitle

Creativity in data visualization is the ability to ask new and appropriate questions and find new and appropriate representations with data. One way that learners can engage with data visualization is through visualization activities: activities that are \textit{hands-on tasks with the goal of learning, reflecting, discussing, or designing data visualisations}~\cite{VisActivities20}. There are many benefits to an active learning strategy. It encourages people to think critically about the problem, helps retain information better, encourages people to take risks and spark new creative ideas, and in a group setting, it develops collaborative skills. 

\begin{figure}
    \vspace{-3mm}\hspace{-2mm}\includegraphics[width=1.05\columnwidth]{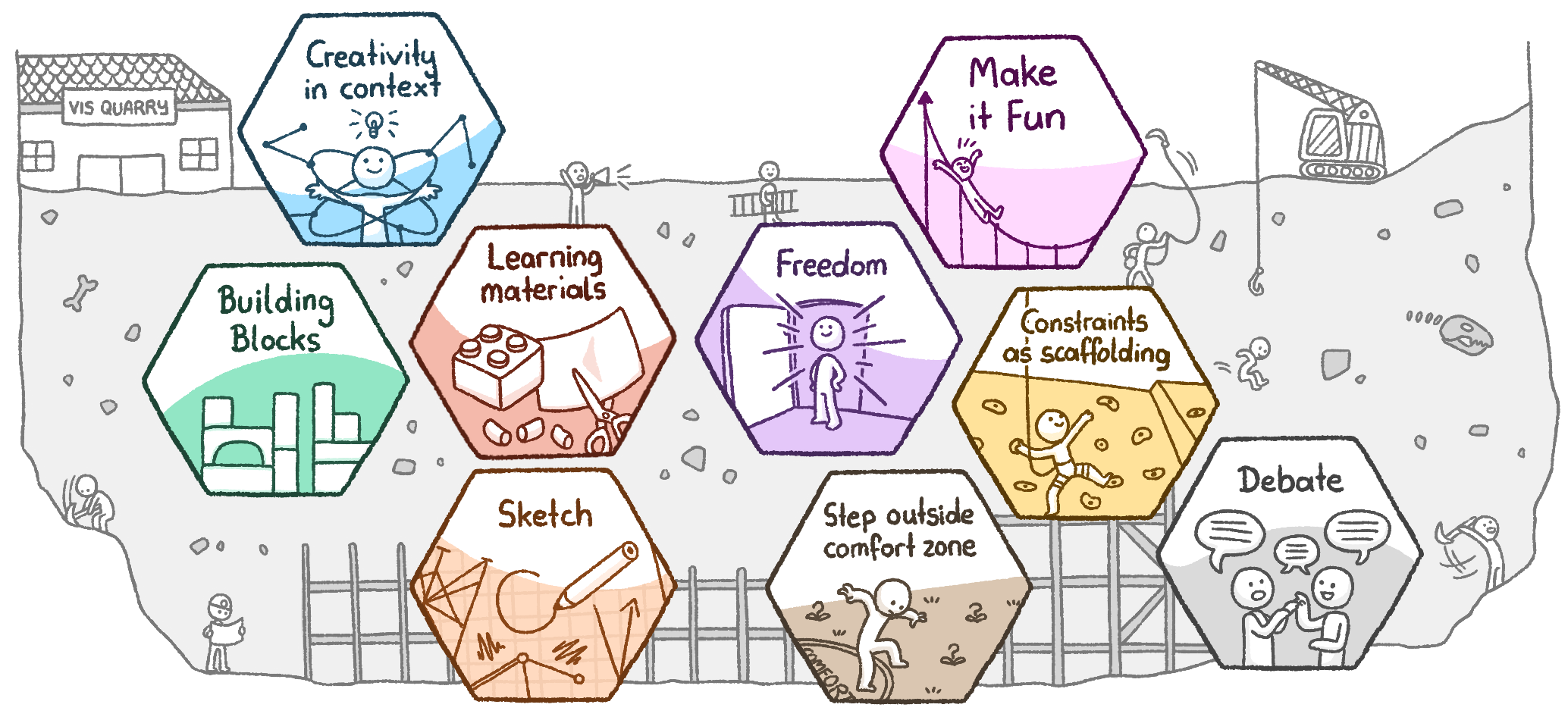}
    \vspace{-3mm}
    \caption{Nine reflections on creative visualization learning activities.}
    \label{fig:reflectionsPict} \vspace{-2mm}
\end{figure}

Recently, we have seen an increasing interest for, and discussions of visualisation activities, with several workshops~\cite{VisActivities20,VisActivities21} and a special edition~\cite{BachETALSpecialIssue2021} on the topic. On reflection, as authors we started to contemplate what lessons can be learned, and whether we can develop guidelines. The activities discussed in this paper have been created by visualization researchers and designers, and applied to a broad range of contexts, and for different audiences. Some involve sketching, others could be to locate and scavenge for different visualization examples, while others could be physical. Activities typically involve a series of steps both for participants and for instructors. They can involve analogue tools such as pen and paper~\cite{WalnySketching}, everyday physical materials~\cite{huron2017let}, or digital visualization tools (e.g., Tableau, D3.js), and the duration of activities can vary from a few minutes to several weeks. Whatever the type or style of activity, the ideas is the same: get people to think about the concepts in a creative and playful way. But, what can we learn? Are there general principles of good practice? Can we create some guides?

This paper reflects on how we design, run, and think about activities. We present a group of nine considerations (\autoref{fig:reflectionsPict}). These represent a  snapshot of ideas; a representative sample. Through them, we hope to enable discussion within the community, and provide a starting point for people to consider what are best practices and lessons learned. Our long term goal is to support the community in developing best practices and guidelines that will engender creative thinking. We want to challenge educators to think differently, and use creative activities in their lessons. We want to stretch learners to approach learning in a variety of ways. After discussing the background and methods, we present the nine strategies for creative visualization activities (\autoref{sec:building_blocks}-\ref{sec:show-them}), followed by discussion.

\section{Background and Methods}
The paper was initiated from discussions at the 2022 Schloss Dagstuhl seminar ``Visualization Empowerment: How to teach and learn''~\cite{DagVisEmp2022}, and motivated from discussions at workshops on VisActivities~\cite{VisActivities20,VisActivities21} which discussed education and activities in depth. Getting together to discuss creativity in visualization, and enthusing about all things creative in visualization, we started to share our experiences, challenges and techniques.  We realised that while collectively some of our ideas were similar, others were new. In our teaching we were using different computing technologies and languages, and a variety of resources, books, and activities. In the Dagstuhl seminar we used Miro boards, Google documents, shared photographs, websites and other materials. We were learning from each other's experiences, knowledge and techniques. We got thinking, perhaps we could adapt an activity from someone in the group for our own teaching. Or better still, our reflections and considerations could become a set of guides for others to adopt.  

We (the authors) are all educators and researchers in data visualization. We come from broad and diverse backgrounds. Individually, we teach to both undergraduate and postgraduate students, and also deliver courses to the public, ranging from small (10) to big (150) class sizes. While most of the students we teach to have backgrounds in computing, data science, human computer interaction and engineering predominantly, we are all starting to teach to those with broader skills, and some of the group taught at Design schools, and to students in Art, Biomedical Engineering, Design, Economics, Mathematics, Statistics, Medicine and Information Systems. 
The breadth and variety of educators at the seminar was further evident when we discussed our teaching practices, materials used, assessment methods --- collectively we covered the full range of coursework, graded exercises, and end of course examination --- and activities. By discussing and summarising our backgrounds, we set different tasks and got learners to implement solutions using different tools; including Processing, JavaScript, R/ggplot, or Tableau. Through this discussion, we realised that by choosing one of these tools, we (as the teacher) were developing different skills in the students. For instance, Processing encourages people to break down the solution into marks on the page, whereas Tableau encourages people to think at a higher level of abstraction.

Together, we discussed visualization activities such as the Five Design-sheet method for visualization design~\cite{roberts2015sketching}, workshops for data comics and storytelling~\cite{bach2018design}, methods for design immersion~\cite{Hall}, design spaces and patterns to practice visualization design~\cite{keck2021}, tangible visualization with physical tokens~\cite{VisKit}, sketching exercises for visualizing two quantities~\cite{Ortiz_2012}, gamification~\cite{huynh2020designing, alper2017visualization}, methods for eliciting self-reflection on personal data~\cite{Thudt:2018:SPP:3173574.3173728}, critical thinking sheets~\cite{CriticalThinkingSheets2020}, engaging people with physicalization~\cite{huron2017let, perin2021students}, stealth learning~\cite{burr2021computational},
and activities to improve data literacy~\cite{bhargava2015designing,he2016v}.

When planning and running activities, educators face multiple questions, need to find optimal solutions across many parameters, or need to quickly assess a developing situation \emph{in situ} and react optimally to that situation. For example, planning an activity imposes questions about the (learning) goals, the material and time required, the degree of help provided, the means available for conducting the activity or potentially the need for an assessment. Contextual parameters include the classical constraints of time and space, but also the number of participants, their backgrounds and skills, etc. Eventually, an educator might need to moderate and facilitate the activity while it is running, alone or with the help of other facilitators; they must clearly communicate the activity and goals, guide participants, provide efficient materials, and react to situations that might derail the activity, participant experience, or learning goals. Dealing with these challenges requires experience and tacit knowledge about people, planning, or facilitation. We face these open questions in the context of creativity, and provide strategies to account how educators solve these issues in practices --- which goals they pursue, which decisions they make, what lessons they learnt. 

To draw together the reflections and considerations, we followed a three-stage approach. First, after an initial discussion, we each proposed ideas. Using an affinity diagramming technique we started to organise the ideas. Drawing together ideas that were similar. We named them (gave them an indicative phrase) which helped to summarise the concept, and also helped us to refer back to them. Using a live Google document, we each edited and discussed the ideas together. From these we had a list of over twenty five ideas. Second, we had a broad discussion about the ideas, discussed whether there were more, and if some could be merged. We challenged the ideas, questioning whether there were lessons learned, and how they were useful. We also wanted a concrete example for each, which helped to refine vague ideas into principles. At this stage we collated some together, added more, and deleted others.  Finally, we divided the work into nine main reflections and considerations, and every author chose the strategies they wanted to lead or collaborate with. The quantity nine was chosen arbitrarily. It was a balance between too many and enough. We wanted to have a comprehensive list of ideas to discuss, a representative sample that demonstrates the diversity of backgrounds and teaching experience of the group. \autoref{fig:reflectionsPict} characterises the nine strategies.

We aim to encourage, inspire, and guide other educators with their visualization activities. In the following, each of the nine reflections --- presented in no particular order --- is a personal account by one or a few of the authors. Each reflection starts with a personal experience, discusses challenges and possible solutions, and ends with considerations. Together, these reflections let us conjecture open questions for research and conducting activities.

\section{Introduce composition through concrete and conceptual building blocks}
\label{sec:building_blocks} 

\begin{figure}
    \centering
    \includegraphics[width=\linewidth]{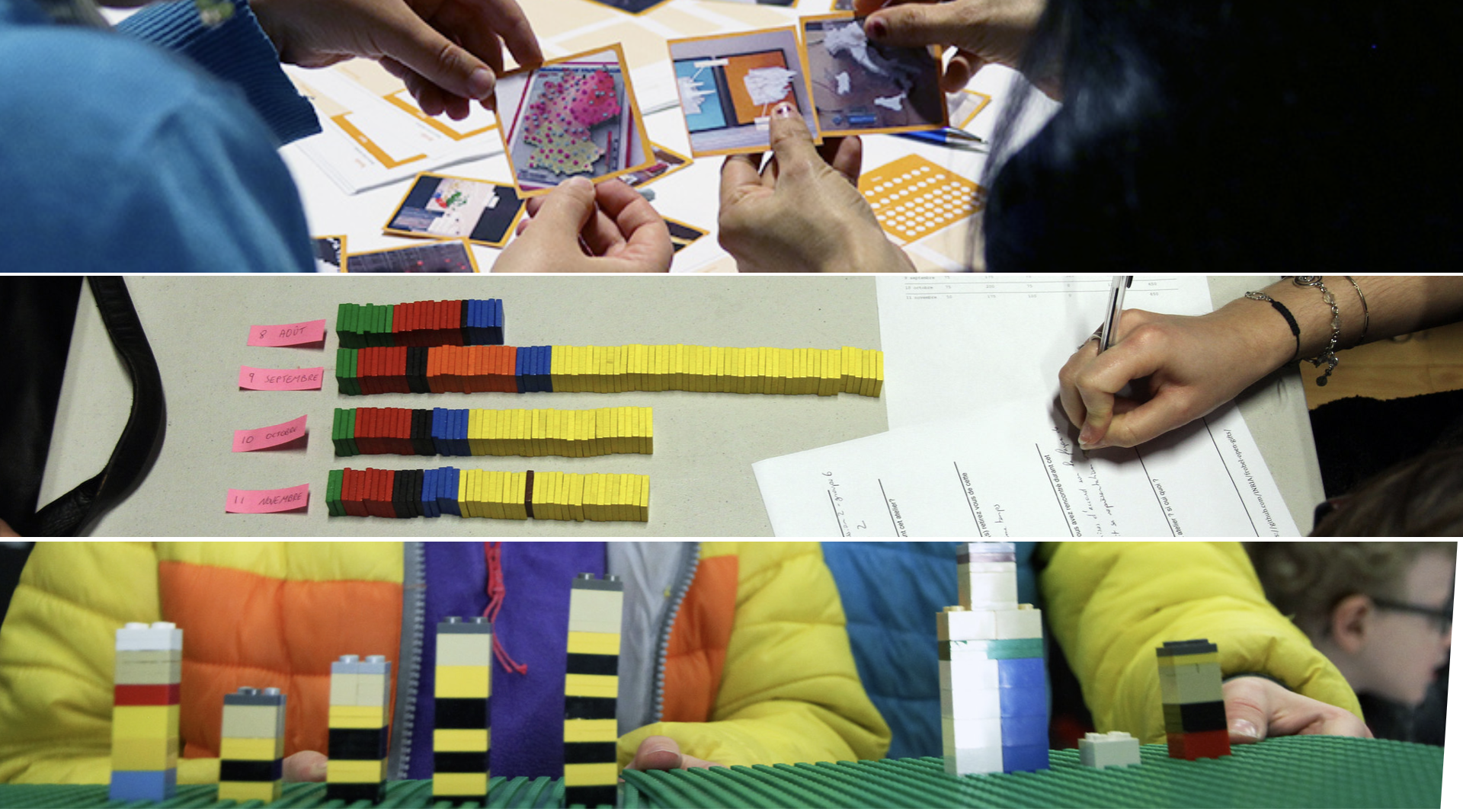}
    \caption{\textbf{Top:} Some participants of ``You named it!'' activity are comparing \href{https://data-physicalisation.github.io/data-phys-cards.html}{data phys cards} to assemble design patterns of physicalization (2018). \textbf{Middle:} Visualization made of wooden tiles with \href{https://www.collectifbam.fr/projets/realisations/phyviz/}{Collectif Bam kit} during Viskit activity at Super Demain (2016). \textbf{Bottom:} Kids showing visualization made out of {\scshape{Lego}} blocks during an activity. (CC BY 2.0 Sylvia Fredriksson).}
    \label{fig:buildingblocks} \vspace{-1mm}
\end{figure}

\textit{Building blocks are simple (conceptual or concrete) elements that can be assembled to make something bigger.} Kids learn counting using their fingers, where each finger represents a value of one. By allocating 10 to a finger, the child can count in tens. Physical building blocks, for kids, have become household names ({\scshape{Lego}}, fidget, kapla, Meccano, etc.), have been used in kindergarten~\cite{leeb1970kindergarten} to teach mathematics~\cite{sarama2004building}, and in K-12 for teaching design and computer programming~\cite{maloney2004scratch}. Abstract and conceptualised building blocks are possible. For instance, Bertin's visual variables are conceptual building blocks for visualization~\cite{Bertin1983}. Moreover, conceptual building blocks could be embedded in card games (see~\autoref{fig:buildingblocks}), to support creativity through simplified construction. 
Building blocks are especially useful for non-visualization experts. People can readily create, update, explain and reflect through their own visualization using simple tangible building blocks in the lab~\cite{huron_tangible_2014} and at home~\cite{Thudt:2018:SPP:3173574.3173728}.

Consequently, researchers have created many pedagogical activities and frameworks (see examples in \autoref{fig:buildingblocks}). People have  used physical or abstract building blocks~\cite{VisKit,willett2016constructive,DataBlokken,Keck20,nolan_teaching_2016} to get learners to construct or even deconstruct information visualization designs. Construction is ``the act of constructing a visualization by assembling blocks, that have previously been assigned a data unit''~\cite{Huron_2014}. Whereas ``the process of identifying and deconstructing a problematic graph and then reconstructing it in a more appropriate form can help students to better interpret, critique, and construct meaningful graphics''~\cite{nolan_teaching_2016}.
As an educator we can start to contemplate what could be used as a building block, what can be constructed and so on. Raw materials such as measures of liquid paint or clay can represent data values~\cite{sturdee2022data-painting}. For example, the length of drinking straws can represent people's heights. Any object or concept that can be considered as individual parts, that can then be constructed. 

\noindent\textbf{Considerations:}
\begin{itemize}[nosep,leftmargin=1em,labelsep=2mm]
  \item The aim of building blocks is to divide the complexity into a multitude of smaller manipulable and modular parts.  
  \item Building blocks have proven to ease and fasten the visualization creation process at the same time as standardizing the results by imposing a structure. 
  \item Building blocks allow to create compelling creative activities. Anything can be used (conceptually and physically). 
  \item They can be used to teach different aspects of visualization, including data, design, mapping, interaction, and so on.

\end{itemize}


\section{Make it Fun}
\label{sec:make-it-fun}
\textit{Feeling good, letting go, enjoying the experience; we are having fun. We are creative when we are having fun. }While as a teacher we cannot force someone to feel good and have \textit{fun},  we can engender a good experience by ourselves relaxing and enjoying the experience~\cite{bisson1996fun}. There is 
a relationship between the student, the teacher and the learning environment, that needs to be developed. In higher education, there is time 
for students to create interpersonal relationships with their teacher~\cite{bernstein-yamashiro_teacher-student_2013}. 

Methods of injecting fun into the college classroom were examined by Tews et al.~\cite{tews_fun_2015} by asking students to rate aspects they enjoyed throughout their learning experience. The highest rated element was humour. 
Adding 
humour into the classroom has a positive impact on creativity and the generation of new ideas~\cite{benjelloun_empirical_2009}. However, it is not always easy to achieve. Studies in an architectural design studio found that when students are in a setting where they are fearful or intimidated by their lecturer or supervisor, their creativity was stunted~\cite{sidawi_impact_2012}. Having a positive class environment encourages a better attitude towards learning and provides students with further motivation in their studies. 

\noindent\textbf{Considerations:}
\begin{itemize}[nosep,leftmargin=1em,labelsep=2mm]
\item Build a fun, trustworthy, and engaging environment where people are `safe to fail'~\cite{Henriksen2021ETAL}. Allow students freedom to explore ideas without the fear of being criticised.
\item Ice breaker activities help to create a relaxed environment, develop personalities, and build foundations for relationships. Perhaps get people to sketch a visualization representing themselves, or `sketch the world'~\cite{SketchingTheWorld}, play games (e.g., Charty Party~\cite{ChartyParty} or the Graphic Continuum ``Match It! Game''~\cite{matchit}).
\end{itemize}


\section{Creativity in Context}
\label{sec:context}
\begin{figure}
    \centering
    \includegraphics[width=\linewidth]{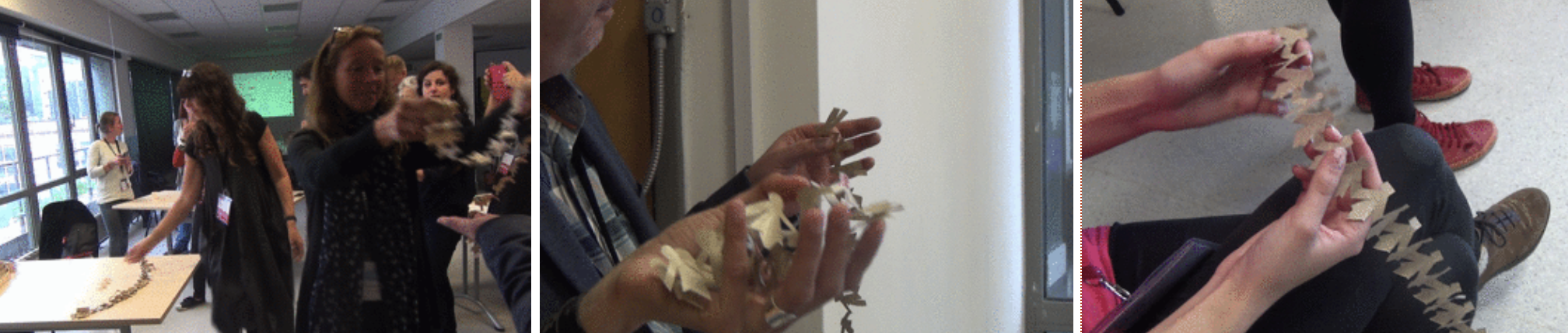}
    \caption{Paper chains of human figures, representing data about asylum seekers.}\vspace{-3mm}
    \label{fig:context}
\end{figure}
\textit{Watching a cooking show inspires you to bake. Attending a craft fair, provides inspirational ideas. Context is important in the creative process.} Context in creative activities is characterized by all the elements surrounding the activity itself. From, say, the creative process itself (e.g., physical \cite{kristensen2004physical}, social\cite{amabile2018creativity}) to the context of the data visualization design process, which includes the context of data production. We can consider any circumstances and settings that surround the core of a problem, or activity, and any parameters and settings that influence it \textit{without being it}. These can be implicit or explicit, known or unknown conditions, and also cultural, situational, spatial, social, political, material, economical, environmental, emotional, institutional, geographical, ideological artifacts and/or phenomena.
Context is an important factor to consider when preparing creative activities. Decisions impact several components of the creative process, from data selection, motivation of creators, the ideation and authoring of the visualization, perception/reception by the public/audiences, to the access to different ideas and resources during various phases of the creation process, including interactions with diverse stakeholders. Through quick activities, students can gain experience with this important relationship between data collection and visualization, for example, by applying methods proposed to assess possibilities for visualization tools where data has not yet been collected~\cite{knudsen2016using}.

\emph{Data is context dependent}. Decisions -- of what is recorded and measured -- impact how we analyze, represent, and interpret information~\cite{boyd2012critical,correll2019ethical,d2020data, dignazio_putting_2019, gitelman2013raw}. Activities that promote data contextualization can help students not only to identify issues with the data, but also to consider how other influencing factors could be visualized, such as ethical aspects. For example, design students were invited to create data-stories about the city of Toronto to position themselves in dialogue with people impacted by this data. While it would be possible to include stakeholders, with larger classes especially, we got students to create data sheets for their data sets~\cite{gebru2021datasheets}. Actively engaging with the social and physical fabric of the city helped students ground their investigations and humanize the data analysis. It promoted understanding contexts that were not present in the data downloaded digitally, encouraging, in many instances, reconsideration of focus and new direction of their data storytelling projects.

In another example, context not only provides relevant data, but engenders an emotive response. In 2015, during a refugee crisis that was happening in the European Union (EU), Huron et al.\ \cite{Huron2017} ran a physicalization creation workshop at the Design Research Society's conference. Two participants presented a paper chain physicalization (\autoref{fig:context}), where each human figure on the paper chain represented an asylum seeker. When presenting their work, they announced emotively ``they are your responsibility now''.  The migration crisis provided the context, which  impacted the entire experience: the organizers’ choice of dataset to share, the participants’ creative process and delivery performance, and finally the public’s reception.

\noindent\textbf{Considerations:}
\begin{itemize}[nosep,leftmargin=1em,labelsep=2mm]
   \item When designing an activity, consider both the contexts of the creative process (e.g., where and when it is taking place, what conditions surround it, social pressures, cultural conditions) and the contexts of the activity itself (e.g., datasets, medium, audiences, expectations).
  \item  Devise activities that are open to accommodate examination of diverse contexts by those involved in the creative process (e.g., social, environmental, political, cultural).
  \item  Participant diversity can benefit the creative process as tasks are examined and explored from different contexts and perspectives.
\end{itemize}

\section{Providing freedom to encourage creativity}%
\label{sec:freedom}

\emph{One way to encourage creativity in learners is to give them freedom in what and how to design and develop their visualization representation ideas. }It is important to allow and foster creativity in learners when teaching how to design and build data visualizations.  
For example, as first-time data visualization instructors, we experimented with a free-form data visualization teaching approach. We used the ``Design Study `Lite' Methodology''~\cite{Borkin2020} to outline the course. While this course was offered for the first time in an engineering department in this format, students from different disciplines or backgrounds could join our data visualization class, pick their own dataset and choice of the target audience, and choose a developing platform to build an innovative data visualization. We collected informal feedback from students throughout the semester about students' experiences as learners. From the collected feedback and personal observations, we share insights and the lessons learned.

Although the course was offered in an engineering department, students from other programs, including Human-Computer Interaction, Data Science, and Medicine, registered for the course. Having students from different disciplines laid the grounds for each individual to offer their unique perspectives during in-class discussions.  
\emph{Flexibility in data selection}. Students were given the option to bring their own data for their course project. Subsequently, they chose a variety of datasets such as results of research projects, performance data at the workplace, and entertainment and leisure-related data. Learners found the ability to choose a dataset of their interest valuable as they could get creative and offer an innovative visualization beyond the already available representations. 

\emph{Flexibility in audience selection}. By having the option to choose the audience of data visualization, students said they could target their visualization design for audiences beyond their classmates. For example, one student chose to design a data visualization representing all air flight accidents over the past 100 years for people who have a flying phobia. Their goal was to show the infrequency of flight accidents. Other students mentioned visualizing their research data empowered them to better articulate their idea and communicate important insights about their data to their supervisors and peers.

\emph{Flexibility in selecting tools} for creating the visualization systems ensured that students could learn and apply visualization techniques regardless of their programming knowledge. Students chose tools such as Tableau storyboard, JavaScript and D3, and Python based on their knowledge and skills. Thus, instead of focusing on learning how to code, students could focus on developing creative data visualizations. However, it can be difficult to introduce various developing platforms in class to support all students' choices. 

\noindent\textbf{Considerations:}
\begin{itemize}[nosep,leftmargin=1em,labelsep=2mm]
\item A student centred approach, where students can make individual choices, can build students' ownership on the task.

\item When students see different approaches and solutions from their peers, it can enrich in-class discussions, they can learn from each other, which can spark creativity. 
\item Allowing flexibility in target audience, choice of data, and tool selection, opens opportunities to design more innovative visualizations. 
\item When offering these freedoms, the instructors' added workload needs to be addressed and supported, particularly when evaluating learners.
\end{itemize}


\section{Different learning materials facilitate learning}
\label{sec:material} 
\emph{Learning requires complex and often uneven development steps. Using different materials can facilitate learning and support the creative process, which is crucial when designing data visualizations}. 
Besides the main-stream digital resources used when teaching visual activities, such as a computer, phone, or tablet, different materials like cardboard, paper, pens, plastics, {\rmfamily\scshape Lego} bricks~\cite{zenk_supporting_2021}, cards~\cite{viscards_2022,matchit,huron2017let}, other tangible objects, or even people can be leveraged to adhere to individual learning styles and understand data visualization processes from different perspectives. 
Methods like these are also often used in creativity and innovation workshops (cf. design thinking)~\cite{zenk_supporting_2021}. There is also a wide range of instructional methods to draw upon, such as active learning~\cite{bonwell-activeLearning-1991}, deconstruct/reconstruct methods~\cite{nolan_teaching_2016}, experience-based learning, storytelling, comics~\cite{matuk_how_2021} and many more. For example, multimedia presentation encourages active cognitive processing and promotes meaningful learning~\cite{MAYER2003125}.

There are many benefits to using different materials. Findings by Chen et al.~\cite{chen_influence_2020} show that variety of material properties (colors, densities, masses, lengths, shapes, and so on) of physical objects facilitate the creative process. Especially in visualization education, students can be often hindered in their creativity by the vast possibilities of digital visualization tools in the early stage of the design process. Designing with {\rmfamily\scshape Lego} lowers this barrier, and students have been shown to be more willing to make changes to their early designs, and consequently were able to create final designs of high visual quality~\cite{ranscombe_designing_2020}. 

There are barriers and challenges. Some students want to dive right into the raw data and improve their coding skills, neglecting the creative process~\cite{roberts2015sketching}. While some students are reluctant to use materials that they perceive as unconventional.
In classes on problem solving in the design process of data visualizations, introducing sketching activities (such as using the Five Design-Sheets method~\cite{roberts2015sketching,roberts2017five} or creating data comics~\cite{wang_teaching_2019}) enables students to process their data through multiple creative channels, gaining insights about their data by reflecting on their sketches with their peers.

\noindent\textbf{Considerations:}
\begin{itemize}[nosep,leftmargin=1em,labelsep=2mm]
    \item To ensure that they are successful, a respectful and patient learning environment is needed, where students can be encouraged to try new methods and embrace creativity in their thinking.
    \item  The fruitfulness of such methods can highly depend on student group dynamics. Therefore, some time might have to be spent on e.g., mitigating students' fear of creating `bad' drawings, being open to critical discussion, etc. 
    \item It is also important to personally feel comfortable with the teaching practices as an instructor.
\end{itemize}


\section{Debate and Discussions} 
\label{sec:debate} 
In the context of participatory visualization activities, \emph{the debate and discussion of visualization constructions, best practices, and methodologies might spark curiosity and engage participants} on learning of visualization concepts, methods, and open problems.

Debates and discussions constitute a lively part of academic visualization research, which can be brought into the classroom. Examples include  ChartJunk~\cite{few2011chartjunk,borkin2013makes}, use of a rainbow colormap~\cite{borland2007rainbow}, 2D versus 3D visualizations, benefits of pie charts, to biased visualizations~\cite{szafir2018good}. Many examples have been captured at workshops such as \href{www.vislies.org}{VisLies} and AltVis at IEEE VIS, where alternative visualization work can be shared.
Moreover, debates have been proposed as a strategy for learning in several disciplines such as political, social, and education sciences~\cite{oros2007let}. They work as a strategic vehicle to engage students by questioning prescriptive knowledge, analyzing opposite viewpoints, and exercising critical thinking. 
The process does not require students to have a specific background or previous knowledge. Instead participants can engage on research around the debate and consolidate their own ideas as well as share and contrast \emph{diversity} of viewpoints, that at the end can increase creativity and spark curiosity to learn more about the topics at hand.

Sometimes learning new concepts may seem tough or perceived as being boring by students.
Peer-based discussions have shown positive impact among the participants, during the lectures, and also asynchronously through online discussions as shown by previous teaching experiences~\cite{diehl2021visguided}. 
For example, during the last three years we have exercised debate and discussions at the University of Zürich with the support of the VisGuides\cite{diehl2018visguides} discussion platform. The platform enables people to pose questions and discuss ideas. It forms a large collected body of evidence of different opinions. 
Among the benefits observed and surveyed after the course from the participants were: more participation during the lectures, exercise of critical thinking, interest to dive deeper into the topics, and awareness of the current visualization challenges and open problems.

\noindent\textbf{Considerations:}
\begin{itemize}[nosep,leftmargin=1em,labelsep=2mm]
  \item Before engaging in a debate of visualization concepts or guidelines, tools and training should be provided to participants such as instructions on general debate protocols, content analysis, and clear goals for the debate or discussion.
  \item Debate and discussions should be supported by evidence. We consider evidence not only peer reviewed publications but also other informal publications on websites, blogs, and forums, as long as they contribute to the discussion.
  \item The outcome of debates or discussions might be open ended. For instance: open challenges, new scenarios of applications, or can provide closure to a problem: evidence of applications, limitations, and clarifications, for example. 
  
\end{itemize}
\section{The Need for a Sketch}
\label{sec:sketch} 
\emph{Sketching supports a creative design process, externalizes ideas, records ideas in a way that can be shared, helps to confirm good (or bad) ideas, and can form part of an assessed submission process.}

To illustrate the challenge, without sketches students struggled to explore choices. In our example, students with a background in computer science and mathematics were tasked to create a dashboard on air-quality data, measured over several years in different cities. Their dashboard was required to demonstrate different visualizations that interacted with each other. They could decide on which data characteristics to display, individual visualization types, and what interaction to provide (cf. freedom/choices in \autoref{sec:freedom} and comfort-zone, \autoref{sec:comfort-zone}). But many students found it difficult to decide between alternative visualisation techniques, whether a line chart, scatterplot or perhaps a geographic visualization. They asked questions, such as ``how should the visualisations be arranged'', ``what interaction process would be best for a user'', ``how to incorporate temporal or spatial data?'' With being of a (business-) computer science and mathematics background, they focused on analytical aspects. Sketching made a huge difference. Even when they used simple, very draft sketches, they conceived, discussed and constantly improved their design ideas. They sized, positioned different elements, tried color combinations, and saw the bigger picture in order to develop and improve a visualization design.

Bonnici et al.~\cite{Bonnici2019SketchbasedIA} showed that when participants were allowed to sketch -- during various experiments with applications in the visualization domain -- they had less difficulty, and achieved better functionality in their results. In terms of data-input Walny et al.~\cite{WalnySketching} showed that participants relating sketches to datasets and representing the data in pictorial drawings had deeper observations and questions about the data they were exploring. There are several ways to use sketching in design, such as the Five Design-Sheets method~\cite{roberts2017five}, online sketching \cite{Browne2011DataAO}, workshops for data comics and storytelling~\cite{bach2018design}, and critical thinking sheets~\cite{CriticalThinkingSheets2020}.

\noindent\textbf{Considerations:}
\begin{itemize}[nosep,leftmargin=1em,labelsep=2mm]
\item Students should be aware that there is no right and wrong in their sketched designs. The goal (if any design design structures are followed, such as the Five Design-Sheets) is to support their creative process.
\item Sketches created without a computer~\cite{Roberts2017ExperienceTeaching} particularly helps students with an affinity for programming to concentrate on the design process and to push implementation into the background.
\item Encourage students to talk about and discuss their sketches. This can also help to develop sketches further.
\end{itemize}


\section{Step Outside Your Comfort Zone as a Catalyst for Creativity}
\label{sec:comfort-zone} 
We often hear, or rather see as a Venn diagram, that ``where the magic happens" is well outside of ``your comfort zone". A strategy to support development of creative thinking skills in learners include activities that \emph{force the learner out of their comfort zone, in terms of tools they are allowed to use, instructions they can rely on, resources they are provided with, and constraints they have to satisfy}. From our experience, one of the main obstacles to creativity in visualization can be attributed to the urge to turn a given spreadsheet of data into a visual representation of these data, using traditional computational tools such as Tableau or D3---perhaps because people believe this is what is expected, or maybe they do not know other ways. But this approach is severely limiting as more often than not, the resulting visualization is (i) heavily driven by the only available data in the spreadsheet, and is (ii) constrained by the learner's technical skills as well as the expressive scope of the visualization tool.

One way to move learners out of their comfort zone is to challenge their status quo by shifting the parameters under which they are asked to approach visualization design. Remove the traditional reliance and anchoring on predefined expectations of a typical data science course. One way to do so is through an exercise to ``envision the ideal visualization about a topic of your choice''. Learners have to assume that they have access to the best possible data they need to communicate their message and are encouraged to mock ideal data tables from scratch. They are required to execute their vision by sketching on paper, without computers. These activities may build or benefit of previous reflections and considerations, such as Freedom (\autoref{sec:freedom}). For instance, by lifting limitations associated to availability and quality of data, tool capabilities, and technical abilities, should help ignite creativity. However, these techniques can be insufficient as a mechanism to encourage creativity, as many students tend to fall back into familiar chart and map designs --- that's what they know best! Here, learners benefit from being imposed yet another set of parameters, through revised instructions or by drawing design constraint flashcards, which require them to rethink their solution entirely and therefore stimulate creative fluids. For instance, learners can be told that ``color is not allowed: communicate the same message without color'', or be given a set color palette, or different target audience and medium (e.g., a mobile phone app for domain experts versus a poster for non-experts).

\noindent\textbf{Considerations:}
\begin{itemize}[nosep,leftmargin=1em,labelsep=2mm]
\item Giving learners full agency over the choice of topic that aligns with their values can result in greater engagement and build learners' resilience through sense of purpose~\cite{waxman2003review}, which results in improved learning.  
\item While removing data and tool constraints expands the possibilities for learners to explore what is otherwise not possible, such freedom can result in learners feeling at lost. Consider providing learners with resources for the more disoriented ones to get their footing, e.g., suggest seed ideas or theme such as ``visualization for social good'', and a list of data repositories.
\item Many students are intimidated by sketching. To mitigate  low confidence in drawing skills, consider complementary activities where students can practice sketching while also appreciating that conveying the idea does not necessarily require professional drawing skills, in a fun way~\cite{roberts2015sketching}. 
\end{itemize}


\section{Constraints as scaffolding}
\label{sec:show-them} 
\emph{Constraints as scaffolding can be good. They can be useful, and help people be more creative.} For example, asking students to write a short story from a blank page can be difficult. But instead, asking someone to write a story with them as a lead character, and with a superpower that they lose at age 18, is another matter. The outline gives structure, sets a specific mindset and allows the participant to kick into the story. It can help engender creative thought, because it removes some of the unknowns from the creative process. When faced with too much choice, people can be uncomfortable and not know how to proceed. Faced with a blank sheet of paper, and working on a design challenge, people can be frightened to make the first mark~\cite{roberts2017five}. 

The process, though, needs to be fully explained. Teaching visualization to undergraduate students, we use the Five Design-Sheet method, which splits the creative challenge into five steps each with five parts~\cite{roberts2015sketching}. As a solution, we play a sped-up recording of the full five-sheets creative process. It helps to explain the process, by providing a recorded example of someone doing one.

There are different aspects of the creative process that can be scaffolded. We have already seen a few examples in different sections. For instance, we could constrain the materials that are used to create the visualisation (\autoref{sec:material}), or stipulate that certain colors are used, which perhaps pushes people out of their comfort zone (\autoref{sec:comfort-zone}). In another example, using the Explanatory Visualization Framework~\cite{RobertsETAL2018EVF}, we scaffold the whole creative assessment, breaking it down into steps (cf. building blocks, \autoref{sec:building_blocks}). Other scaffolds could be considered, such to focus on rewards or dreams of the student, or get people to enter into a public competition~\cite{MAKSIC2021100835}. 

\noindent\textbf{Considerations:}
\begin{itemize}[nosep,leftmargin=1em,labelsep=2mm]
\item Think what structures can be used in the creative process. 
\item Make sure the scaffolding is clearly explained (consider using videos to explain the process).
\end{itemize}


\section{Discussion}
We have only scratched the surface of possible considerations. Our ideas represent a snapshot, and small part of the outcomes from the Dagstuhl seminar. They are a starting point for creating a broader set of visualization guidelines, for creativity in teaching. They provide ideas, experiences and strategies that other people can follow. Drawing all these ideas together what is the take home message? What overarching considerations should people appreciate?

\textbf{Be balanced in your choices.}
We have too many choices. The landscape of options at our disposal is rich: from sketching on paper, using tangible objects, to programming of data-to-visual mappings allowing to plot and re-plot millions of data items automatically. Different strategies afford a certain type of activities, audiences, and resulting learning objectives. Consequently, educators need to make decisions over their learning outcome, tasks, content, structure of the unit, depending on their situation. Decisions need to be tempered by the quantity of time we (as educators) have allocated, how many students there are on the course, if there is lab assistant, and so on.

\textbf{Be selective.}
We are not proposing for educators to follow everything. Teachers, learners, researchers or designers sit in different contexts and must decide what is suitable for their situation. It will not be possible for one person to integrate all nine ideas into their teaching. Nor is it feasible to expect students to learn everything in one course. Visualization, as a science and as a practice, is very broad. It encompasses many facets, from principles stemming from empirical work and theories in cognition, perception, information science, and design; to methods and techniques adopted and adapted from psychology, sociology, and human-computer interaction; to technical skills in the fields of computer science and engineering. 

\textbf{Choosing a tool.}
The question about which visualization authoring methods and tools as strategies to support learning is a often a tricky one for instructors when designing activities and assessment for a given audience. 
Choosing the right programming language, is one of the major decisions that defines how visualization design and creativity is taught. Computational solutions such as D3, R/ggplot or Python are extremely powerful in that they support both reproducibility and expressiveness, but they also require to integrate programming instruction to teach these languages, or setting stringent pre-requisites, as the learning curve is steep. Light-weight computational solutions include more approachable wrapping libraries such as Observable~\cite{observable}, Vega~\cite{satyanarayan2016vega}, or Processing~\cite{processing}, allowing to lower the threshold for novice learners, while supporting stealth development of programming skills~\cite{burr2021computational}. Programming provides more flexibility and freedom than template-based tools such as Tableau, but still require learners to be computing-savvy, and/or instructors to devote part of the curriculum to programming instruction. Ideally, teachers should strive for authoring strategies with both a \textit{low threshold}~\cite{myers2000past}, i.e., methods that do not require or require very little prior knowledge or experience; and a \textit{high ceiling}~\cite{myers2000past}, i.e., tools that allow  visualization creators to achieve efficiency (through replication, automation) and expressiveness (through flexibility in design). Whatever strategy used, the teacher needs to feel comfortable with the idea, knowledgeable of the material, and confident of the approach.

\textbf{Non-computer tools.}
In our reflections, we stress the value of using non-computational tools such as sketching~(\autoref{sec:sketch} and \ref{sec:comfort-zone}), and using building blocks~(\autoref{sec:building_blocks}), which are easy to manipulate and do not require prior training, while also being fun and playful~(\autoref{sec:make-it-fun}). These are not only excellent options to help learners unleash their creativity (anything is possible in sketching!), they also are highly valuable teaching interventions allowing learners to focus on the visualization principles themselves rather than the practical challenges of executing an ambitious vision. 

Teachers should also be considerate of technical skills acquired. Low-tech approaches such as drawing and sketching may not scale up, and might have limited buy-in in some professional settings. Students interested in learning about visualization often expect, and are eager to learn how to use computing libraries and toolkits that support data manipulations and automated rendering through defining data-to-visual mappings. In our experience, we find it is difficult to achieve both deep technical training (i.e., teach JavaScript+D3, or R+ggplot+Shiny) and visualization science (i.e., cognition, perception, theory, creativity, storytelling, etc.) within a typical self-contained course spanning a dozen classes. Current offerings required the instructor to either sacrifice creative and critical thinking activities to support programming instruction, or sacrifice technical training or restrict access to a tech-savvy audience.
While visualization researchers continue to push for authoring methods and tools that further bridge the gap between free-form creative sketching and powerful, reproducible data-to-visual mappings~\cite{satyanarayan2019critical}, e.g., exploring approaches such as lazy data binding~\cite{liu2018data, xia2018dataink, zhang2020dataquilt, kim2016data}, we are still lacking a clear solution that is both suitable to a broad audience, is valued and utilized in industry, and supports teaching of the full range of learning outcomes in the area.

\textbf{Consider the learner, learner journey, and learner outcomes.}
Teaching principled approaches to visualizations design involves the teacher engaging with students in \emph{applying} concepts learned~\cite{anderson2001taxonomy}. While this is true for any module, students who (for whatever reason) are reluctant to be in the class (perhaps as it is a requirement of their course), can be less-engaged, and reluctant to consider alternative and creative teaching practices. 
But the challenges that are faced in Higher Education and adult learning, occur at all levels. Past studies have outlined the lack of concrete standards for visualization as early as in elementary-level curricula~\cite{chevalier2018observations}. The lack of clear instructional goals and learning objectives can be an important contributing factor to observed poor visualization literacy in the general population~\cite{borner2016investigating}. Critically, visualization instructors are left with little resources to set learning objectives, besides bearing inspiration from colleagues' past course offerings, or building from visualization textbooks (such as Munzner's~\cite{munzner2014visualization}) but in general learning objectives, exercises, and activities, are missing from these resources. Activities can be used as an agent of inclusion and help to empower marginalized groups. 

Finally, we hope this paper contributes as an additional source for inspiration, but stress for the need for the community to eventually outline and consolidate a standard set of high-level learning objectives to guide curricula development.


\vspace{-1mm}\acknowledgments{
This work was funded by BMK under the ICT of the Future program via the SEVA project (no. 874018), by the Austrian Science Fund as part of the Vis4Schools project (I 5622-N), by the GFF NÖ as part of the dissertation project VisOn (SC18-008) and VisToon (SC20 - 014), by French government funding  managed by the National Research Agency under the 
Investments for the Future program (PIA) grant ANR-21-
ESRE-0030 (CONTINUUM), and by Canadian NSERC funds (RGPIN-2018-05072) and (RGPIN-2021-04222). 
We thank Prof. Tatiana von Landesberger for her contributions to the discussions, and the reviewers for their comments and suggestions.}

\bibliographystyle{abbrv-doi}
\bibliography{creativeVisLearnActivities}
\end{document}